\documentstyle[12pt,eqsecnum,aps,floats,epsf]{revtex}

\begin{document}

\begin{flushright}
CEBAF-TH-96-09 \\
DOE/ER/40561-258-INT96-19-02 \\ 
\end{flushright}

\begin{center}
{\Large \bf 
Interaction of small size\\
wave packet with hadron target}  
\end{center}
\begin{center}
{L. FRANKFURT\footnotemark 
\footnotetext{\noindent On leave of
    absence from the St.Petersburg Nuclear Physics Institute, 
Russian Federation
    }\\
{\em School of Physics and Astronomy, }\\ 
{\em Raymond and Beverly Sackler Faculty of Exact Sciences,}\\
{\em Tel Aviv University, 69978 
Tel Aviv, Israel}\\ [2.0mm] 
A. RADYUSHKIN\footnotemark 
\footnotetext{\noindent Also  Laboratory of
Theoretical Physics, JINR, Dubna, Russian Federation} \\
{\em Physics Department, Old Dominion University, 
Norfolk, VA 23529,USA} \\
{\em and}\\ {\em Thomas Jefferson National Accelerator Facility,} \\
 {\em Newport News, VA 23606, USA}\\ [2.0mm] 
M. STRIKMAN\footnotemark 
\footnotetext{\noindent Also  St.Petersburg Nuclear
 Physics Institute, Russian Federation} \\
{\em Department of Physics,
Pennsylvania State University, }\\
{\em University Park, PA  16802, USA} \\ }
\end{center}

\begin{abstract}

We calculate in QCD the cross section for the 
scattering of an  energetic small-size wave
packet off a hadron target.
We use our  results to  study  the small-$\sigma$
behaviour of $P_{\pi N}(\sigma)$,
the distribution over cross section 
for the pion-nucleon scattering, in the
leading $\alpha_{s}$-order.

\end{abstract}
PACS numbers:  12.38 Bx ;  13.60.-r ; 13.60.Hb

\newpage

\section{Introduction}

Recently phenomena involving interactions of 
hadrons in small-size
configurations
have been 
intensively
discussed both in relation with the phenomenon of
color transparency and vector meson electroproduction 
observed at HERA energies.
There is also a deep relation between presence of the weakly interacting
small size configurations in hadrons and phenomenon of cross section
fluctuations in the interactions of the hadrons which manifests itself
in the inelastic coherent diffraction processes: 
$h +N(A) \rightarrow X +N(A)$, see Ref.\cite{fms}.
In this paper, we focus on the systematic derivation of the formulae
for the interaction of the color singlet 
$q \bar q$ pair  having a small transverse size
with a hadron target. Then, we use these formulae to calculate 
the probability of the distribution for the interaction of a photon and 
a pion with a target for small interaction cross sections.
Although some of equations deduced in the paper  existed before 
no derivations with analysis of 
their  accuracy has been presented.

The paper is organized as follows. In Section 2, we consider the
 virtual forward Compton amplitude  in the small-$x$
region where it is dominated by the photon-gluon scattering
subprocess. We outline there a 
derivation of the  basic formula  expressing the 
total cross section $\sigma_{\gamma^* T}$ 
as a convolution of the gluon distribution amplitude
$G_T(x,Q^2)$ and the 
$\gamma g$
scattering cross section.
In Section 3, we write  down 
the $\gamma g$ cross section
 in terms 
of the $\bar qq$ light-cone wave functions of the virtual photons.
In the next section, we calculate the cross section distribution
$P_{\gamma^*}(\sigma)$ for the virtual photon.
In Section 5, we discuss the quark-hadron 
duality interplay between the perturbative free-quark results
and  contributions due to low-lying resonances.
Finally, in Section 6, we calculate the cross section distribution
for the pion $P_{\pi}(\sigma)$ in the small cross section limit
where it is governed by $\bar qq$ configurations
having small spatial size. 
Basing on QCD evolution equation we evaluate also the functional dependence of 
$P_{\pi}(\sigma)\rightarrow 0$ on $\sigma$ and on 
the incident 
energy.

\section{Hard $\gamma^{*}T$ total cross section and the interaction of
small size configurations.}

\setcounter{equation}{0}

\begin{figure}[b]
\mbox{
   \epsfxsize=12cm
 \epsfysize=5cm
 \hspace{2cm}  
  \epsffile{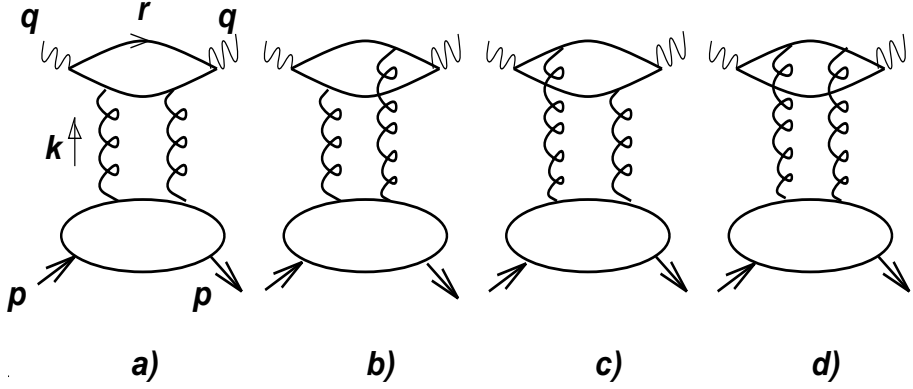}  }
  \vspace{0.5cm}
{\caption{\label{fig:1}
Leading small-$x$ contribution to the 
forward virtual Compton amplitude. 
   }}
\end{figure}

        Let us consider a particular contribution
into the $\gamma^{*}T$ cross section corresponding to a transformation 
of the virtual photon  when $\gamma^{*}$  converts into a
$Q \bar{Q}$ pair with 
quarks having a large relative transverse momentum.  
Usually, this 
contribution is  written 
as a convolution of the  infinite momentum 
frame wave function of the 
target with the pQCD calculable coefficient function 
describing the short-distance   propagation
of the particles between two virtual photon vertices.    
Our aim is to express the relevant coefficient function
 in terms  of the light-cone wave functions of
the virtual photons as viewed from the  reference 
frame where the target is at rest.
The contribution we 
are interested in  is given by the sum of 
diagrams shown in Fig.\ref{fig:1}.

The  lower  blob corresponds to  the gluon distribution in the target. 
It is convenient to parameterize the gluon momentum $k$ 
in  terms of the Sudakov variables
\begin{equation}
k = - \alpha q^{\prime} + \beta p^{\prime} + k_t \,  
\ \  \  \ \ \ d^{4}k = \frac{ s}2 d \alpha  d \beta  d^2 k_t .
\label{eq:1.1} \end{equation}
Here $q^{\prime} $ and $p^{\prime} $ are light-like 
momenta  related  to $p,q$ by
\begin{equation}
q=  q^{\prime} +\frac{q^2}{2 (p^{\prime} q^{\prime})}p^{\prime} \  \  ; \ 
\  p=  p^{\prime}  +\frac{p^2}{2 (p^{\prime}  q^{\prime} )} q^{\prime}  ;
 \label{eq:1.2} \end{equation}
\begin{equation}
  2 (pq) = 2 (p^{\prime}  q^{\prime} )+ 
\frac{q^2 p^2}{2 (p^{\prime}  q^{\prime} )}  \ .
\label{eq:1.3} \end{equation}
For our  goals, the most interesting region
is that  of small values of the Bjorken parameter: $x =
{-q^2}/{2(pq)} \rightarrow 0$, where 
we may safely approximate   $2 (p^{\prime}  q^{\prime})=s$.

In the $d^{4} k$ integral,
the region $k_t^2 \sim  Q^2$ corresponds to 
the next-order $\alpha_s$ correction,  
so we will take into 
account only  the contribution of the region $k_t^2
\ll Q^2$.  This corresponds to  the leading $\alpha_{s} \log
 Q^2/\Lambda^2 $ approximation in which the
 ${\cal O} (\alpha_{s})$ corrections  are 
neglected.  In this kinematical region,  the contribution of
the diagrams shown in Fig. 1 can be considerably simplified. 
The essential region of integration is
\begin{equation}
\mid k^2 \mid = \mid \alpha \beta s + k_t^2 \mid \ll
Q^2 .
\label{eq:1.4} \end{equation}
Note that  $\beta s \sim Q^2 \propto  (mass)^2$ 
of the $q \bar {q}$ state produced 
by
$\gamma^{*}$.  Hence, 
\begin{equation}
\alpha \ll 1
\label{eq:1.5} \end{equation}
is the  essential  region of
integration  over $\alpha$. 
It is convenient to write   the  propagator 
$ d_{\mu \bar \mu }(k)/{k^2}$ of  the exchanged
gluon in the light-cone 
 gauge $q^{\prime}_{\mu} A^{\mu} =0$,
in which 
$$ d_{\mu \bar \mu }(k) =  - g_{\mu \bar \mu } +
{{q^{\prime}_{\mu}  k_{\bar \mu} + k_{\mu}  q^{\prime}_{\bar \mu}}
\over{(k q^{\prime})}}.$$
It  can be shown ($cf.$ \cite{Gribov})  that in our case 
the  $k_{\mu}  q^{\prime}_{\bar \mu}$  part
of the propagator dominates.   Using 
eq.(\ref{eq:1.5}), the dominant part of the gluon
propagator can be further simplified:
\begin{equation}
d_{\mu \bar \mu }(k) \simeq \frac{p_{\mu}^{\prime} 
q_{\lambda}^{\prime} }{ ( p^{\prime}  q^{\prime} )}.
\label{eq:1.6}
 \end{equation}

In other words,  it is sufficient to take into account 
only the longitudinal polarization of  the exchanged gluons.
 Indeed, let us estimate the contribution 
due to  exchange of a transversely polarized gluon:
\begin{equation}
\delta \sigma \sim {1\over 
(2\pi)^4}
\int {T^{\gamma^*}_{\mu_{\perp}\lambda_{\perp}}\over s} d  (\beta s)
\,\,\, \int
{T^{T}_{{\mu}_{\perp}{\lambda}_{\perp}}\over s}  d
(\alpha s )\frac{d^2 k_t}{(k^2)^2}.
\label{eq:1.7} \end{equation}
Here,  $T^{(\gamma^*)}_{\mu_{\perp \lambda_{\perp}}}$ is the
imaginary part of the amplitude of the
$\gamma^{*}$ scattering off a gluon
given by the lowest-order Feynman diagrams 
and  
$T^{T}_{{\mu}_{\perp}{\lambda}_{\perp}}$ is that
for the gluon scattering off a target $T$.  
Using the   fact that, at high energies,
the  Feynman amplitude of processes due to exchange by 
two elementary fermions tends to constant 
\cite {Azimov} we obtain:
$$\int T^{\gamma^*}_{\mu_{\perp}\lambda_{\perp}}d (\beta s) 
\propto  {(\beta s)^{2}\over Q^2} \simeq {(Q^2)}.$$ 
In this estimate we use also scaling over $Q^2$ in the box diagram.
The amplitude 
due to the exchange by two vector
particles increases like $s$, and we have 
$$\int  T^{(T)}_{\mu_{\perp}\lambda_{\perp}} 
d \alpha s 
\propto {(\alpha s)^{2+n}}.$$
Here  $n > 0$, since, according to  the QCD evolution equations,
the deep inelastic amplitudes increase with 
energy in region of 
applicability of the perturbative QCD.

As a result of this power counting estimate we obtain:
\begin{eqnarray}
\delta \sigma \sim \int \frac{(\beta s)^2}{s} \frac{(\alpha s)^{2+n}}{s}
\frac{d^2k_t}{(\alpha \beta s + k_t^2)^2} \theta
(k_t^2 < k^2_{t0}) \theta (\alpha \beta s \leq
k^2_{t0}) \nonumber \\ \approx \int
\theta(k^2_{t} < k^2_{t0}) 
{(\alpha s)^{n}}
\propto (\alpha s)^{n} .
\label{eq:1.8}
\end{eqnarray}
Here, we  substituted $\alpha \beta s \sim k_t^2$.  
Thus, due to the presence of the 
factor $\alpha$ in eq.(\ref{eq:1.8}), 
in the leading $\alpha_s \log Q^2/ \lambda^2$ approximation,
  the contribution due to the exchange of a transversely polarized 
 gluon is negligible  compared
to the contribution of the 
longitudinal  polarization specified by eq.(\ref{eq:1.6}).
We use here the observation that in 
QCD
the power $n$ characterizing the energy dependence of the amplitude is the same
for scattering of transversely and longitudinally polarized gluons. 

Using the gluon propagator in the form given by eq.(\ref{eq:1.6}),
we  get the following expression for the total contribution
 of the diagrams shown in Fig. 1:
\begin{equation}
{\rm Im}  \,    M= \int \frac{d^{4}k}{(2 \pi)^{4}i}\frac{1}{(k ^2)^2} 2
\cdot {\rm Im}  \,    T^{a b (P)}_{\mu \lambda} {\rm Im}  \,   
T^{a b (T)}_{\bar \mu
\bar  \lambda} d_{\mu \bar \mu}(k)  
d_{\lambda \bar  \lambda}(k)
\label{eq:1.9} \end{equation}
Here $T^{a b (P)}_{\mu \lambda} = T^{a b}_{\mu
\lambda}(\gamma ^{*} g \rightarrow Q \bar{Q})$ is the sum of the box diagrams
describing the $\gamma^{*}g$ scattering and  $T^{a b (T)}_{\bar \mu
\bar \lambda}$ is the amplitude of the gluon scattering off the target $T$. 

        Using the dominance of the 
longitudinal gluon polarization (\ref{eq:1.6}) and  
incorporating
 eq.(\ref{eq:1.1})
we can rewrite  eq.(\ref{eq:1.9}) as
\begin{equation}
\frac{{\rm Im}  \,    M}{s}= \int \frac{s d \alpha  d \beta  d^2k_t}{2 (2
\pi)^{4} (k^2)^2} \frac{2 {\rm Im}  \,    T^{a b (P)}_{\mu
\lambda}p_{\mu}p_{\lambda}}{4 (pq)^2}
\cdot \frac{4{\rm Im}  \,     T^{a
b(T)}_{\bar{\mu}\bar{\lambda}}q_{\bar{\mu}}
q_{\bar{\lambda}}}{s}.
\label{eq:1.11} \end{equation}
 
Now we will use 
 the fact that
\begin{equation}
T^{a b (P)}_{\mu \lambda} k_{\mu} =
T^{a b (P)}_{\mu \lambda} k_{\lambda}=0, 
\label{eq:1.10} \end{equation}
 since the box diagram contains no gluons and,  therefore, the  
Ward identities
in this approximation are the same 
as in an  Abelian gauge theory
($a, b$ are the color indices).
From eq.(\ref{eq:1.1})  and (\ref{eq:1.10}) it follows  that
\begin{equation}
\frac{{\rm Im}  \,    T^{a b(P)}_{\mu \lambda} 
p_{\mu}p_{\lambda}}{4(pq)^2}
= \frac{{\rm Im}  \,    T^{a b
(P)}_{\mu_{\perp}\lambda_{\perp}}k_{t}^{\mu}k_{t}^{\lambda}}{(\beta
s)^2}
\label{eq:1.12} \end{equation}
and 
\begin{equation}
\frac{{\rm Im}  \,    M}{s} = \int d \beta s \frac{\frac{1}2 \sum T^{a b
(P)}_
{\mu_{\perp}\mu_{\perp}}}
{(\beta s)^2} \int \frac{d \alpha s
d^2k_t}{(2\pi)^{4}(k^2)^2} 
\frac{4k_t^2{\rm Im}  \,
T^{a b
(T)}_{\bar{\mu}\bar{\lambda}}q_{\bar \mu}q_{\bar \lambda}}{s^2}.
\label{eq:1.13} \end{equation}
It is useful to define the cross section of $\gamma^{*}$ scattering off
a gluon $g$ averaged over the gluon color:
\begin{equation}
\delta_{a b}\cdot s^{\prime}  \cdot \sigma (\gamma^{*}g \rightarrow q
\bar{q}) = \frac{1}2 \sum_{\mu_{\perp}=1,2} {\rm Im}  \,    T^{a b
(P)}_{\mu_{\perp}\mu_{\perp}},
\label{eq:1.15} \end{equation}
where $s^{\prime} $ is the invariant mass of 
the produced $q \bar{q}$ system: $s^{\prime}  
=(k + q)^2 \simeq \beta s -Q^2$.
  Thus,
\begin{equation}
\sigma_{\gamma^{*}T}= \frac{{\rm Im}  \,    M}{s} = \int \frac{d \beta }{\beta}
\sigma (\gamma^{*} g \rightarrow \bar{q}q) \int \frac{s d \alpha 
d^2k_t}{(2\pi)^{4}(k^2)^2} k_t^2 \sum_{a} 
\frac{4 {\rm Im}  \,    T^{a(T)}_{\bar{\mu}\bar{\lambda}}
q_{\bar \mu}q_{\bar \lambda}
}{s^2}.
\label{eq:1.16} \end{equation}
In the leading $\alpha_{s} \log Q^2$ approximation,  
we can  substitute $k^2$ by $k_t^2$.
Comparing  our result  with  the QCD-improved
parton model expression for the production of heavy quarks 
(see $e.g.,$ \cite{hq}),
we observe that 
\begin{equation}
 \int \frac{s d \alpha  d^2k_t}{(2 \pi)^{4}
k_t^2} \sum_{a} 4 {\rm Im}  \,     T^{a a
(T)}_{\bar{\mu}\bar{\lambda}}
q_{\bar{\mu}}q_{\bar{\lambda}} =  \beta G_T(\beta, Q^2),
\label{eq:1.17} \end{equation}
where $G_T$ is the gluon distribution in a target $T$.  
This gives 
\begin{equation}
\sigma_{\gamma^{*}T} = \int \sigma_{\gamma^{*}g} \frac{d \beta }
{\beta} 
[\beta G_T (\beta,Q^2)].
\label{eq:1.18} \end{equation}
The first argument of 
$ G(\beta,Q^2)$ is 
$\beta = {Q^2+M^2\over s} $.
Here  $M$  is the mass of the  produced $q \bar q$ pair,
which is typically of the order  of  $Q$. 
Hence,  the essential region of integration is $\beta  \sim x$.
As the  evolution scale for the gluon
distribution function, we take $Q^2$. 
Of course,    higher-order $\alpha_{s}$ corrections 
 may change $Q^2$ by some
numerical factor.  This scale-fixing ambiguity  
is a usual feature of  the leading $\alpha_{s}
\log Q^2$-approximation.

\section{Light-cone wave functions and $\sigma_{\gamma^{*}g}$}

Now let us express $\sigma_{\gamma^{*}g}$
in terms of  the light-cone wave functions of the virtual photon.  
To this end, we write  down 
the four-momenta   $r_{1}(r_2)$ of quark (antiquark) 
in the box  in terms of the light-cone variables  
${r}_{1}= \{r_1^+,r_1^-, r_t \}$ with
$r_1^+ = \eta q^+$ and take  the integral over $r_1^-$ 
by residue.  Introducing the lowest-order perturbative
$\bar q q$ light-cone wave functions 
of the virtual photon \cite{BrodLep}
\begin{equation}
\psi_{\mu} = \frac{\bar {U} (r_{1}) \gamma_{\mu}
 U(-r_2)}{\frac{m^2+r_{t}^2}{\eta(1-\eta)}+Q^2}
\frac{1}{\sqrt{\eta (1- \eta)}},
\label{eq:1.20} \end{equation}
we obtain the following expression for the sum of the box diagrams:
\begin{eqnarray} 
\lefteqn{\hspace{-10cm} \int d \alpha s \frac{{\rm Im}  \,     
T^{a b (p)}_{\mu
\lambda}p_{\mu}p_{\lambda}}{s^2} =} 
\nonumber \\ e^2g_s^2 \int \frac{d \eta
d^2r_{t}}{2(2 \pi)^{3}} \pi \psi_{\mu} (\eta, r_{t}) 
\{ 2 \psi_{\mu} (\eta, r_{t}) - \psi_{\mu}
(\eta, r_{t} + k_t) - \psi_{\mu} (\eta,r_{t}-k_t) \} F_{a}F_{b} ,
\label{eq:1.19} 
\end{eqnarray}
where   $g^2_{s}$ is the QCD coupling constant and   
 $F_{a} = \frac{\lambda_{a}}2$,  $\lambda_{a}$ 
being  the Gell-Mann matrices of the $SU(3)_{c}$ group in the 
fundamental representation.

It is convenient to rewrite this formula in the  impact
parameter space:
\begin{equation}
\psi_{\mu} (x,r_{t}) = \int \psi_{\mu}(x,b)
e^{ir_{t} b}d^2b .
\label{eq:1.21} \end{equation}
Then 
\begin{equation}
\int d \alpha s \frac{{\rm Im}  \,     T^{P}_{\mu \lambda} 
p^{\mu}p^{\lambda}}{(2pq)^2}
= \int \psi^2_{\mu} (x,b) \frac{dxd^2b}{4 \pi}g^2_{s} 
\left \{ \pi
[2-e^{ik_tb}- e^{-ik_tb}]
Tr (F_{a}F_{b}) \right \}.
\label{eq:1.22} \end{equation}
Within the leading $\alpha_{s} \log Q^2$ approximation, 
to obtain eq.(\ref{eq:1.12}),
it is necessary to decompose exponent into a power series over $(k_tb)$
and to keep terms up to the second order in $k^2_{t}$.
 Combining eqs. (\ref{eq:1.21}), (\ref{eq:1.11}), 
and (\ref{eq:1.17}), we obtain
\begin{equation}
\sigma_{\gamma^{*}T}=e^2 \int \psi^2_{\mu}(\eta, b) \frac{dzd^2b}{4
\pi}N_{c} \left \{ \frac{1}{N_{c}} g^2_{s} \pi
\frac{(k_tb)^2}{k_t^2} Tr\frac{F^2}{8} \right \} \time \cdot G_T(x,
\lambda
/b^2).
\label{eq:1.23} 
\end{equation}
Here, factor $\lambda$ can be estimated from analysis of 
$\sigma_L(\gamma^*N)$ cross section. 
Since the gluon density increases when $x$ decreases,
 $\lambda$ slowly increases with decrease of $x$
\cite{FKS}.  For $x \sim 10^{-3}$,  $\lambda \approx 9$.

It is instructive  to represent  $\sigma_{\gamma^{*}T}$ in the form 
\begin{equation}
\sigma_{\gamma^{*}T} = e^2 \int \psi^2_{\mu} (\eta, b) \frac{d \eta
d^2b}{4 \pi} N_{c} \cdot \sigma_{t}^{q \bar{q}}(b^2).
\label{eq:1.25} 
\end{equation}
Here, $\sigma_{t}^{q \bar{q}}$ is the 
cross section for the interaction of a colorless small transverse
size $q \bar{q}$-pair with the target $T$:
\begin{equation}
\sigma_{t}^{q \bar{q}} = g^2_{s} \pi \frac{b^2}2
\frac{1}{N_{c}} Tr \left ( \frac{F^2}{8} \right ) x G_T(x, 
\lambda
/b^2).
\label{eq:1.24} 
\end{equation}
 This expression was obtained originally in \cite{BBFS93,FMS93}.
 As usual, $N_{c}$ is the
number of colors,  and
the Casimir operator of the $SU(3)$ group 
in  the fundamental representation
can be easily calculated:
\begin{equation}
\frac{1}{8} \frac{1}{N_{c}} Tr F^2 = \frac{1}{3} Tr F^2_{3} =
\frac{1}{6} .
\label{eq:1.26} 
\end{equation}
Combining all the numbers together, we finally obtain:
\begin{equation}
\sigma^{q \bar{q}}_{T} = \frac{\pi^2}{3} b^2 \left[ x
G_T(x, \lambda/b^2) \right] \alpha_{s}(\lambda/b^2).
\label{eq:1.27}
 \end{equation}
Here $b =
(b_{q}-b_{\bar{q}})$.
 This formula describes the
essence of the color transparency (CT) phenomenon
($cf$. discussion in \cite{BFGMS}): $q\bar{q}$ configuration of 
a small spatial size has a small interaction  cross section.
However, for  sufficiently small $x$,  
the interaction becomes strong due to
the formation of the soft gluon field.  
In this respect,  eq.(\ref{eq:1.23}) predicts
the  interaction of a small  size
configuration  which is qualitatively different  from that
of the  models 
of F. Low \cite{low}  and J. Gunion and 
D.Soper \cite{gunionsoper}.  The fact that 
$\sigma^{q \bar{q}}_{T} $ is proportional to 
the gluon distribution 
in eq.(\ref{eq:1.24}) increasing in the small-$x$ 
region, has important experimental consequences,
$e.g.$ it makes it  possible  to 
observe the  small-size quark configurations at HERA in the
electroproduction of vector mesons at small $x$.
In fact, eq.(\ref{eq:1.23}) can 
be inferred from a formula derived in  ref.
\cite{mu}  within a model approximation to QCD. 
Using  some simple tricks, 
one can  also obtain eq.(\ref{eq:1.23}) 
from a  formula obtained in \cite{levin}  
within the leading $\alpha_s \ln x$
approximation of $QCD$  combined with 
some  
bold
assumptions concerning  the  parton model structure.

Using (\ref{eq:1.19}), we can 
calculate distribution over
cross section for the fast photon or pion projectible for small
$\sigma$ ($cf.$\cite{BBFS93}).

\section{Distribution of $P_{\gamma^* N}(\sigma)$ for the photon
projectile.}
\setcounter{equation}{0}

In the previous section we have derived  
eq.(\ref{eq:1.23}) which expresses
the $\sigma_{\gamma^{*}T}$ cross section in terms of the light-cone wave
functions of the virtual photon $\gamma^{*}$.  
This formula gives us the  possibility to
calculate another useful quantity - distribution over cross section

$P_{\gamma^{*} T}(\sigma)$. 
By definition, the differential probability  that the 
virtual photon $\gamma^{*}$ interacts with the target
$T$ with the cross section $\sigma$. 
In other words, 
the  experimentally observable 
 total cross section in terms of  $P(\sigma)$ is given by 
\begin{equation}
\sigma_{\gamma^{*}N} = \int P_{\gamma^{*} N} (\sigma) \sigma d \sigma.
\label{eq:2.1} 
\end{equation}

In refs. \cite {FP,GW}, it has been suggested to represent 
the cross section $\sigma$ in terms of the 
eigenstates of the $S$-matrix.
In the case of small $\sigma$, as a result 
of color screening and asymptotic freedom, 
the scattering state is a $q \bar q$ pair.
So, the  contribution of small  $\sigma$
has the form of eq.(\ref{eq:1.25}).
Using eq.(\ref{eq:1.24}), we can write:
\begin{equation}
e\sigma_{\gamma^{*}N} = e^2 \int 
\psi^2_{\gamma^{*}} (\eta, b) \frac{d \nu}{4 
\pi}N_{c}
\sigma \frac{\pi db^2}{d \sigma} d \sigma \, .
\label{eq:2.2} 
\end{equation}
Let us  define 
\begin{equation}
P_{\gamma^{*} N}(\sigma \rightarrow 0) = \int e^2
\psi^2_{\gamma^{*}}(\nu,b) \frac{d \nu}{4} N_{c} \frac{\pi db^2}{d
\sigma},
\label{eq:2.3} 
\end{equation}
where $\psi_{\gamma^{*}}(\eta,b)$ is given  by eqs.(\ref{eq:1.20}) and 
(\ref{eq:1.21}). 
It is implied here that the functional dependence of  
$b$ on  $\sigma$ in eq.(\ref{eq:2.3})
should be calculated from  eq.(\ref{eq:1.24}).
Now, we can rewrite eq.(\ref{eq:2.2})  in the form of eq.(\ref{eq:2.1}).
Though our derivation is applicable for the interactions with
small $\sigma$, eq.(\ref{eq:2.1}) 
has a more general nature. 
In fact,
it has been understood long ago 
\cite{FP,GW,PumplinMiettinen} that many 
features of
the interaction of a fast projectile can be described in terms of
distribution over cross section.  An important advantage of such a 
quantity is
that it accurately takes into account  diffractive processes.
Some 
properties of $P(\sigma)$ have been discussed in detail in \cite 
{Heiselberg}. 
However, for our purposes, it is sufficient to
consider  $P_{\gamma^{*}N}(\sigma)$ 
in the limit 
 $\sigma \rightarrow 0$.

  In general, eq. (\ref{eq:2.3})
 predicts a rather involved dependence of $P(\sigma)$
on $\ln \sigma$ at small $\sigma$. However, this dependence 
can be easily calculated
using QCD evolution equations. 
The distinctive feature of eq.(\ref{eq:2.3}) is that
\begin{equation}
P_{\gamma^{*} N} (\sigma \rightarrow 0) \mid_{\sigma \rightarrow 0} \sim
\frac{1}{\sigma}\;\;\;\;\;\;\;\;\; \mbox{up to}\; \log
(\sigma / \sigma_{0}) \  \mbox{terms}.
\label{eq:2.4} \end{equation}

\setcounter{equation}{0}

\section{Transition to mesons}

The perturbative version of the virtual photon wave function 
$\psi_{\mu}(\eta, r_t) $ (3.1) can be written 
through a dispersion integral
\begin{equation}
\psi_{\mu}(\eta, r_t)  = \frac1{\pi} \int_0^{\infty} 
\psi_{\mu}^{\bar q q} (\kappa; \eta, r_t )
\frac{d\kappa^2}{\kappa^2+Q^2} 
\label{eq:disp}
 \end{equation}
where 
\begin{equation}
\psi_{\mu}^{\bar q q} ( \kappa; \eta, r_t) =  \frac{\bar {U} (\eta q_+) 
\gamma_{\mu}
 U((1-\eta)q_+) } {\sqrt{\eta (1- \eta)}}  
\delta \left (\kappa^2- {\frac{m_q^2+r_t^2}{\eta(1-\eta)}} \right ) 
\label{eq:3.2} \end{equation}
is the wave function of a non-interacting 
$\bar q q$-pair with invariant mass 
$\kappa$. 
The  interaction between the quarks modifies 
the virtual photon wave function 
$\psi_{\mu}(\eta, r_t) \to \Psi_{\mu}(\eta, r_t)$,
and  the dispersion representation 
\begin{equation}
\psi_{\mu}(\eta, r_t)  = \frac1{\pi} \int_0^{\infty} 
\psi_{\mu}^{\bar q q} (\kappa; \eta, r_t )
\frac{d\kappa^2}{\kappa^2+Q^2} 
\label{eq:disp2}
 \end{equation}
for the
``exact'' wave function $\Psi_{\mu}(\eta, r_t)$ 
is in terms of the 
modified spectral density $\psi_{\mu}^{hadr} (\kappa; \eta,r_t)$
in which, 
 instead of the free-quark approximation
$\psi_{\mu}^{\bar q q} ( \kappa ; \eta, r_t)$, one has 
a sum over  resonances,  the $\rho$-meson being the dominant feature
in  the low-$\kappa$ region:
\begin{equation}
\psi_{\mu}^{\bar q q} (\kappa ; \eta, r_t) 
\to \psi_{\mu}^{hadr} (\kappa; \eta,r_t)
= g_{\rho} \psi_{\mu}^{\rho} (\eta, r_t) 
\delta(\kappa^2 -m_{\rho}^2) + 
\psi_{\mu}^{higher \, states} (\kappa; \eta, r_t) 
\label{eq:3.3} \end{equation}
where  $g_{\rho}$ is  the 
magnitude of the $\rho$-state projection 
onto the electromagnetic current.
 At large $\kappa$, the resonances are  wide, and their sum 
rapidly approaches the free-quark value, $i.e.,$  
one has a perfect quark-hadron 
duality\footnote{Note, that since the 
large-$\kappa$ behaviour of  
$\psi_{\mu}^{hadr} (\kappa; \eta,r_t)$
coincides with that of 
$\psi_{\mu}^{\bar q q} ( \kappa ; \eta, r_t)$,
the dispersion integral
in eq.(\ref{eq:disp2}) has the same convergence properties 
as that in eq.(\ref{eq:disp}), $i.e.$, 
no subtractions are needed in eq.(\ref{eq:disp2}).}.
For sufficiently large $Q^2$,
the dispersion integral (\ref{eq:disp2})
is dominated by higher states, and 
the  free-quark approximation  is completely justified.
Decreasing $Q^2$, one would observe mismatch 
between the free-quark calculation and the dispersion integral 
over the resonances. Such a situation  is well known 
from QCD sum rules:  the difference between the resonance
and free-quark spectra  is described by  
power corrections $(1/Q^2)^N$.    The  usual procedure 
is to approximate the 
higher states by the free-quark contribution  
(``first resonance plus continuum''  model) 
$$ \psi_{\mu}^{higher \, states} (\kappa; \eta, r_t) = 
\theta(\kappa^2 > s^{\rho}_0)  \psi_{\mu}^{\bar q q} (\eta, r_t)
$$
where $ s^{\rho}_0$ is the effective threshold for higher resonances
in the $\rho$-channel and then fix its value by the requirement 
of the best agreement between the two sides of the resulting sum rule
\begin{equation}
\frac1{\pi} \int_0^{s^{\rho}_0} 
\left ( \pi g_{\rho} \psi_{\mu}^{\rho} (\eta, r_t) 
\delta(\kappa^2 -m_{\rho}^2) - 
\psi_{\mu}^{\bar q q} (\kappa; \eta, r_t) \right )
\frac{d \kappa^2}{ \kappa^2 +Q^2} 
= \sum_{N=2} \frac{A_N}{(Q^2)^N}.
\end{equation}
After fixing  $ s^{\rho}_0$ from the magnitude of the 
power corrections ${A_N}/{(Q^2)^N}$, one can take 
the limit $Q^2 \to \infty$
to get the local duality relation 
\begin{equation}
\pi g_{\rho} \psi_{\mu}^{\rho} (\eta, r_t) =  \int_0^{s^{\rho}_0}
\psi_{\mu}^{\bar q q} (\kappa; \eta, r_t) \, d \kappa^2.
\end{equation}  
In other words,   the $\rho$-meson wave function in such an approach
 is dual  to the free-quark wave functions 
integrated over the duality interval
$0 \leq \kappa^2 \leq   s^{\rho}_0$.

For the forward virtual Compton amplitude, 
the dispersion  representation can be  applied  both for the 
initial and ``final''  virtual photon. 
However, taking only the $\rho$-meson contribution 
in the dispersion  integral for the final state,
one naturally obtains  the amplitude  for the 
$\gamma^* T \to \rho T$ transition considered in ref.  \cite{BFGMS}. 
Furthermore, picking out  
the $\rho$-meson contribution 
in both dispersion  integrals  one would get the amplitude
for the $ \rho T \to \rho T$ scattering.
This idea   can be also used  to study 
the pion diffractive electroproduction and the
pion diffractive scattering.

\section{Calculation of $P_{\pi N}(\sigma \rightarrow 0)$.}
\setcounter{equation}{0}

To  analyze the pion scattering, we  
substitute the electromagnetic current
by the axial current in the original amplitude,
$i.e.,$ simply add $\gamma_5$ in the current vertices.
For massless quarks, the final result  
has the same structure as that for  the vector  current.
Of course,  the $\bar q q$-pair  wave function 
would have an extra $\gamma_5$, and the 
vertex factor analogous to that in eq.(5.2) is 
\begin{equation}
\frac{\bar{U} (x P_+) \gamma^{\mu} \gamma_{5}V((1-x)P_+)}{\sqrt{x(1-x)}}
= P_+^{\mu} \, ,
\label{eq:4.5} \end{equation}
where $P$ is the 4-momentum associated with the axial current. 

 The  projection  of a single-pion state 
onto the axial current is specified by the $\pi \to \mu \nu$
decay constant $f_{\pi}$:
\begin{equation}
\langle 0 \mid J^{A}_{\mu} \mid \pi, P \rangle = \sqrt2
f_{\pi} P_{\mu} \,   .
\label{eq:4.4} \end{equation}
Hence, we should extract  the amplitude  
$\sim P_{\mu}  P_{\nu} $ corresponding to 
the longitudinal polarization of the axial current.
Again, the transition from the virtual amplitude for the currents  
to that involving the pion can be understood 
in terms of the dispersion representation and quark-hadron duality.
In other words, below the effective higher state threshold
$s_0^{\pi}$,  one should substitute the 
free-quark contribution  by that due to the pion pole:
\begin{equation}
 \psi_{5 \mu}^{\bar q q} ( \kappa; \eta,r_t)  \to 
\Psi_{5 \mu}^{hadr} ( \kappa; \eta,r_t) =
q_{\mu} 
\left ( f_{\pi}   \psi_{\pi} (\eta,r_t) 
\delta(\kappa^2 -m_{\pi}^2) + \theta(\kappa^2 > s_0^{\pi}) 
 \psi^{\bar q q}  (\kappa; \eta,r_t) \right ).
        \label{eq:4.0} \end{equation}
The local duality prescription gives a correctly normalized 
wave function provided that $s_0^{\pi} = 16 \pi^2 f_{\pi}^2 \approx 0.67 \, 
GeV^2$. 
Of course, one can use a pion wave function 
different  from that given by the local duality.
However, the duality  considerations justify the
use of the effective two-body wave function (see \cite{apa}).

The actual calculation consists of   the same steps 
as those  leading  to eq.(\ref{eq:1.25}).  
For  a small-size configuration,  we get the following contribution 
$\delta  \sigma_{\pi N}$ into
the scattering cross section: 
\begin{equation}
\delta  \sigma_{\pi N}= \int | \psi_{\pi} (\eta, b)|^2 \frac{d \eta 
d^2b}{4 \pi} N_{c} \sigma^{q \bar{q}}_{N}(b^2) . 
\label{eq:4.1} \end{equation}
Effectively, the vertex $e \psi_{\gamma^{*}}\sqrt{N_{c}}$ is 
substituted by   the pion wave function.
Rewriting $\delta  \sigma_{\pi N}$ as 
\begin{equation}
\delta \sigma_{\pi N} = P_{\pi N}
(\sigma) d \sigma = \frac{db^2}{d \sigma} 
\int |\psi_{\pi} (\eta , b)|^2 \frac{d \eta }{4}
 \sigma d \sigma \ , 
\label{eq:4.2} \end{equation}
 we obtain:
\begin{equation}
P_{\pi N} (\sigma \rightarrow 0) = \frac{db^2}{d \sigma} 
\int |\psi_{\pi} (\eta , b \rightarrow
0)|^2  \frac{d \eta }{4}  \ , 
\label{eq:4.3} \end{equation}
where $\sigma (b^2)$ is given by eq.(\ref{eq:1.27}).

Thus,  $P_{\pi N} (\sigma \rightarrow 0)$  is determined by the pion 
wave function at the origin of the impact parameter space,
or, what is the same, by the integral of the momentum wave function 
$ \psi_{\pi}(\eta, r_t)$ over all transverse momenta $ r_t $. 
This integral formally gives the pion distribution amplitude
$$\varphi_{\pi}(\eta) =  \frac{\sqrt{3}}{(2\pi)^3}
 \int  \psi_{\pi} (\eta, r_t) d^2 r_t .$$
However, in QCD (and in any theory with dimensionless coupling constant),
this integral diverges.  The standard  procedure 
is to supplement the integral with some renormalization  
prescription characterized  by a cut-off parameter $\mu$,
$i.e.,$ $\varphi_{\pi}(\eta) \to \varphi_{\pi}(\eta, \mu)$.
In fact, the Fourier transformation from the momentum to the
impact parameter space
$$ \psi_{\pi}(\eta, b) = \int  \psi_{\pi} (\eta, r_t) 
\frac{d^2 r_t}{(2\pi)^2} $$
for small $b$ can also be treated as a particular
cut-off prescription  with $1/b$ playing the role of the
renormalization parameter $\mu$.  In the $b \to 0$  limit, 
one encounters the singular  $ \log b^2 $ terms. 
It is exactly the  logarithms  which  generate the
evolution of the pion distribution amplitude. 
Summing the logarithms  by  the renormalization group methods gives,
for small $b$:
\begin{equation}
\psi_{\pi} (\eta, b) =  \eta (1- \eta) \sum_{n=0} a_n
C_n^{3/2} (2 \eta -1) \left 
(\frac{ \log b_0^2 \Lambda^2}{\log b^2 
 \Lambda^2} \right )^{\gamma_n/2 \beta_0}
\end{equation}
where $\eta (1- \eta) C_n^{3/2} (2 \eta -1)$ 
are the eigenfunctions of the evolution kernel 
($C_n^{3/2}(2 \eta -1)$ being  the Gegenbauer polynomials), 
the anomalous dimensions $\gamma_n$ are its eigenvalues and 
$\beta_0$ is the one-loop QCD $\beta$-function coefficient. 
The $b_0$-parameter  characterizes the effective
onset of the perturbative evolution.
The coefficients $a_n$ are the Gegenbauer moments 
of the pion wave function at this scale. 
Note that   the anomalous dimension of   the axial 
current vanishes ($\gamma_0 =0$) and all other $\gamma_n$'s are positive.
Hence, after the renormalization  group improvement,
the limit $b \to 0$ is well-defined in this case and 
\begin{equation}
\psi_{\pi} (\eta, b=0) = \sqrt{48}\pi f_{\pi}  \eta (1- \eta),
\label{psiasy}
\end{equation}
where $f_{\pi}=92 \, MeV$.
The absolute normalization of the pion wave
function for $b=0$ is fixed   by the matrix element of the axial current:
\begin{equation}
\int \psi_{\pi} (\eta, r_{t}) \, \frac{d \eta d^2r_{t}}{8 \pi^{3}}=
\frac{f_{\pi}}{\sqrt{N_{c}}} ,
\label{eq:4.7} 
\end{equation}
or in the impact parameter space (see eq.(\ref{eq:1.21}) 
\begin{equation}
\int \psi_{\pi} (\eta, b=0) \, \frac{d \eta}{2 \pi} = 
\frac{f_{\pi}}{\sqrt{N_{c}}}.
\label{eq:4.8} 
\end{equation}

In other words,   for the pion, the singular $\log b$ terms sum into 
harmless $(1/{\log b^2  \Lambda^2})^{\gamma_n/2 \beta_0}$
factors vanishing in the $b \to 0$ limit.
As a result, the 
$\eta$-dependence
of the  pion wave function $\psi_{\pi} (\eta, b)$ in 
the formal $b \to 0$ limit 
always assumes its asymptotic form  $\psi_{\pi} (\eta, b) \sim \eta (1- \eta)$,
irrespectively of its shape 
at the scale $b_0$. 
It is natural to expect that  $b_0$ 
is related to the scale characterizing the 
magnitude of the nonperturbative  momentum distribution
in the pion. The momentum scale  $\mu_0 = \sqrt{s_0^{\pi}} \approx 0.8 \, GeV$ 
suggested by the  local duality is rather large,
and there may exist  a transitional region of 
distances $b \sim b_0$    small 
compared to the pion size but  not small enough
to  produce  sizable   perturbative evolution effects.
In this case, one can try the $\eta$-dependences 
of $\psi_{\pi} (\eta, b)$ different from the asymptotic form.
In fact, the integral 
\begin{equation}
I \equiv \int |\psi_{\pi} (\eta , b)|^2  \frac{d \eta }{4} 
\end{equation}
is  rather insensitive to the evolution effects.
If we take the asymptotic wave
function (\ref{psiasy})
then  
\begin{equation}
P_{\pi} (\sigma \rightarrow 0) =  \frac2{5} \pi^2  f^2_{\pi} 
\frac{db^2}{d \sigma}.
\label{eq:4.12} \end{equation}

Assuming that, at the scale $b=b_0$, the $\eta$-dependence of 
the pion wave function corresponds to the  
Chernyak-Zhitnitsky  \cite{cz} ansatz
\begin{equation}
\psi^{CZ}_{\pi}(\eta,b=b_0) = 5 \sqrt{48} \pi f_{\pi}  \; 
\eta (1 - \eta) (1-2 \eta)^2,
\label{eq:4.11} 
\end{equation}
we obtain 
\begin{equation}
P_{\pi} (\sigma, b=b_0) = \frac{10}{21} 
\pi^2 f^2_{\pi}   \frac{db^2}{d \sigma}.
\label{eq:4.13} \end{equation}
Thus, in this case the evolution would decrease the integral  $I$ 
by $\sim 20$\% when $b$ changes from $b_0$ to $0$. 
 Taking the asymptotic result, we get:
\begin{equation}
P_{\pi} (\sigma \rightarrow 0) = \frac{6}{5}
\frac{f^2_{\pi}}{\alpha_{s}  x G_{N} (x, \lambda/{b^2})}.
\label{eq:4.14} \end{equation}

Distribution
$P_{\pi N} (\sigma)$ was determined in Ref. \cite{BBFS93}
from the
analysis of  the soft diffractive processes for 
$E_{\pi} \approx 200 \, GeV$, see solid curves in Fig.2.
 In the limit $\sigma \ll \left<\sigma\right>$, we 
can compare this result with eq.(\ref{eq:4.13}).
The applicability region of this equation 
is restricted by several  conditions. 
First, $x_{eff}$ should be small enough so that the
average
 longitudinal distances in the scattering process 
$1/2 m_Nx$ 
 are larger than the nucleon size, which corresponds
to $x \alt 0.05$. 
Furthermore, the 
virtualities in the process should be large enough
so 
that
one can apply pQCD which corresponds to the requirement 
$Q^2_{eff} \agt 1 - 2 \, GeV^2$.  In 
our
analysis we also neglect
 the $b$-dependence of the wave function of the $q \bar q$ component
at large $b$ (this is a higher twist effect), 
which restricts consideration
to $b \alt 0.5 fm$. In the 
numerical
calculation, we use the GRV parameterization
\cite{GRV}  since it 
describes well 
the
parton distributions down to $Q^2 \sim 1.5 \, GeV^2$.
 We present results both for 
the
leading 
 and next-to-leading order GRV parameterizations, see dashed curves in Fig.2.
Difference between LO and NLO results  
illustrates range of uncertainties
of the current analysis.
One can see that the results of 
our
calculations are in qualitative agreement with
the phenomenological results of \cite{BBFS93}.

Another interesting feature
of  our results  is a substantial energy dependence of 
$P(\sigma < \left<\sigma \right>) $ on the incident energy due to a fast 
increase of $xG_N(x,Q^2)$ with the decrease of $x$, see Fig.3.
This reflects the fact that the 
probability of point-like configurations in hadrons
decreases with the increase of energy.
Further diffractive data (preferably at higher energies) are necessary
to get better information about $P_{\pi N}(\sigma)$.

        Since the existence of  configurations 
with small spatial size has been 
confirmed experimentally  in the energy dependence and absolute
value of cross section of electroproduction of vector mesons,  we
consider the above result as a reflection of soft matching between
nonperturbative and pQCD regimes.

\section{Summary and conclusions}

In this paper, we applied a pQCD approach to describe 
the basic features of the high-energy interactions
of  a small-size $\bar qq$ configurations  with 
a hadron target. This interaction is proportional
to the gluon distribution function $G_T(x,Q^2)$
 of the target and, hence, the cross 
section is enhanced in the small-$x$
region.  The  $\bar qq$ configuration 
can be described  by the wave functions whose 
particular form is determined by the projection of the 
initial particle ($\gamma^*, \rho$ or $\pi$) 
onto   the $\bar qq$ component.
For small $\sigma$, we calculated the cross section distribution
$P_{\pi}(\sigma)$ for the pion and demonstrated 
that it is rather insensitive
to the specific form of the pion distribution
amplitude.

\section{Acknowledgments}
This work was partially supported by the BSF Grant No. 9200126
and by the US Department of Energy 
under contracts DE-FG02-93ER40771 and  DE-AC05-84ER40150.
Two of us (AR and MS) thank the DOE's Institute for Nuclear Theory
at the University of Washington for its hospitality
and support  during the workshop
``Quark and Gluon  Structure of Nucleons and Nuclei''.

\newpage

\newpage 

\begin{figure}
\mbox{ 
\epsfxsize=15cm
 \epsfysize=15cm
  \epsffile{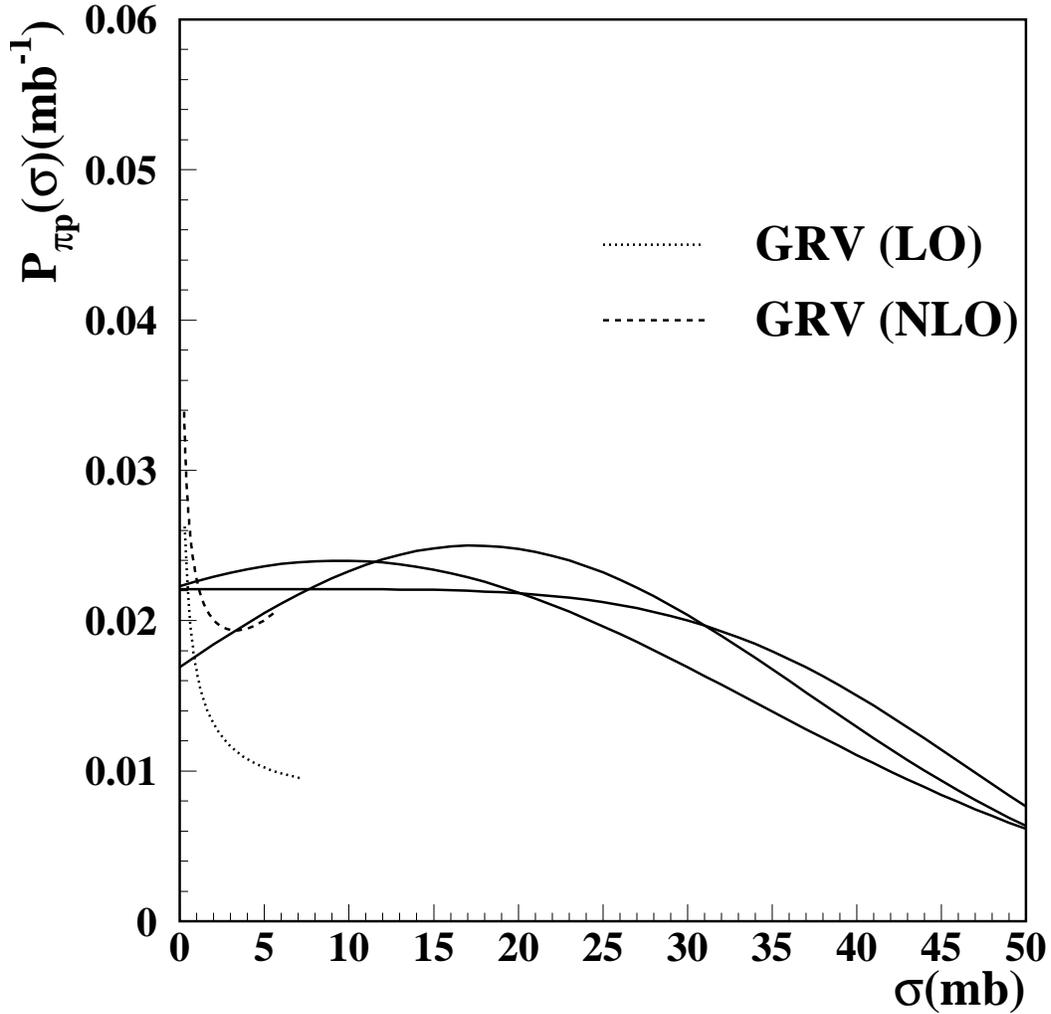}  }
{\caption{\label{fig:2} Comparison of $P_{\pi p}(\sigma) $ calculated in pQCD 
using eq.(\protect\ref{eq:4.13})) and GRV parameterizations%
\protect\cite{GRV} of the gluon density 
 and fits based on the analysis of the soft diffraction
data \protect\cite{BBFS93}.
}}
\end{figure}

\newpage 

\begin{figure}
\mbox{ 
\epsfxsize=15cm
 \epsfysize=15cm
  \epsffile{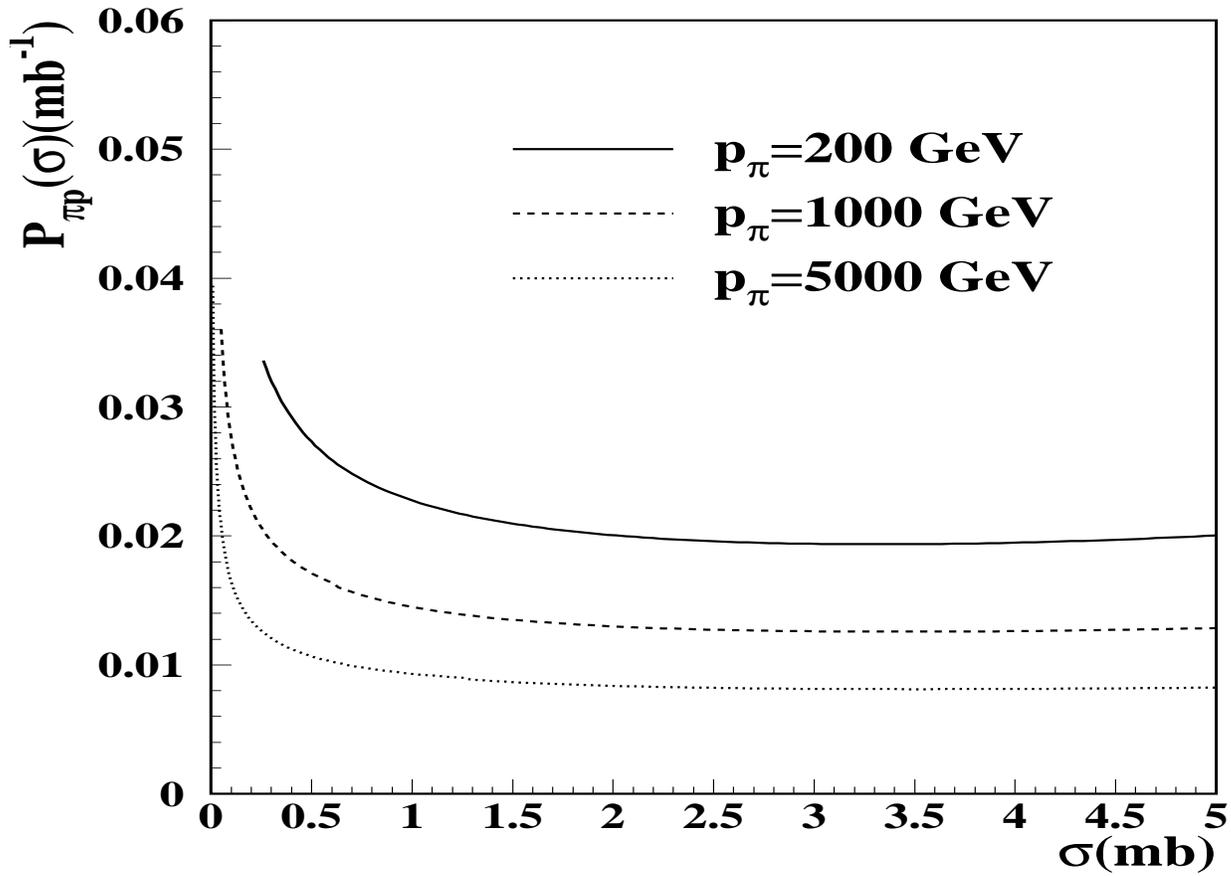}  }
 {\caption{\label{fig:3} Incident momentum dependence of  
$P_{\pi p}(\sigma)$ for 
small $\sigma$ calculated 
using eq.
(\protect\ref{eq:4.13}) and GRV NLO parameterization
\protect\cite{GRV} of the gluon density.} }
\end{figure}

\end{document}